\documentclass[twocolumn,superscriptaddress,prl,10pt,preprintnumbers]{revtex4-2}

\usepackage{amsmath,amssymb,mathtools}
\usepackage{xcolor}
\usepackage{enumerate,enumitem}
\usepackage{bm}

\definecolor{ForestGreen}{rgb}{0.13,0.55,0.13}
\definecolor{burntorange}{rgb}{0.800, 0.333, 0.000}

\newcommand{\bR}{\mathbb{R}}
\newcommand{\bE}{\mathbb{E}}
\newcommand{\cN}{\mathcal{N}}

\newcommand{\cL}{\mathcal{L}}

\usepackage{tensor}

\begin{document}

\title{Gauge Fields in Neural Network Field Theory}

\preprint{INT-PUB-26-041}

\author{Christian Ferko}
\affiliation{Department of Physics, Northeastern University, Boston, MA 02115, USA}
\affiliation{The NSF Institute for Artificial Intelligence and Fundamental Interactions}

\author{Daniel C.\ Hackett}
\affiliation{Institute for Nuclear Theory, University of Washington, Seattle, WA 98195-1550, USA}

\author{James Halverson}
\affiliation{Department of Physics, Northeastern University, Boston, MA 02115, USA}
\affiliation{The NSF Institute for Artificial Intelligence and Fundamental Interactions}

\begin{abstract}

We extend the neural network field theory  correspondence, which has been used to realize scalar fields and fermions as ensembles of random neural networks, to the setting of Abelian gauge fields. Free Maxwell theory is engineered using a vector of single-layer networks with cosine activations and a parameter density which reproduces the Maxwell correlation functions. Gauge invariance and the transformation of the gauge field $A^\mu$ under spacetime rotations are recovered by correlating input and output weights, while retaining statistical independence of neurons. We present numerical simulations using this neural network realization, intrinsically in the continuum rather than on the lattice, finding agreement between simulated observables and exact results.

\end{abstract}

\maketitle

\textbf{Introduction.}---Many quantum field theories (QFTs) are associated with measures on spaces of field configurations, such as the path integral measure. The neural network field theory (NN-FT) correspondence \cite{Halverson:2020trp,Halverson:2021aot} suggests that we instead represent a Euclidean QFT using a probability distribution $P ( \theta )$ over the parameters of a neural network with a prescribed architecture $\phi_\theta ( x )$. Random draws of the parameters $\theta$ then yield random field configurations, and the correlation functions of the theory are computed via parameter-space integrals
\begin{align}\label{scalar_correlator}
    \langle \phi_\theta ( x_1 ) \ldots \phi_\theta ( x_n ) \rangle = \int d \theta \, P ( \theta ) \, \phi_\theta ( x_1 ) \ldots \phi_\theta ( x_n ) \, .
\end{align}
Although the NN-FT correspondence has also been applied to fermions \cite{Frank:2025zuk} and $bc$ ghosts \cite{Frank:2026bui,Ferko:2026kkm}, as the notation in (\ref{scalar_correlator}) suggests, the majority of the NN-FT literature has focused on commuting scalar fields $\phi$. Every scalar theory associated with a non-negative measure admits a NN representation \cite{Ferko:2026axm}, and many cherished aspects of QFT --- such as spacetime and global symmetries \cite{Maiti:2021fpy} and their anomalies \cite{Ferko:2026kkm}, topological features \cite{Ferko:2026ken}, interactions \cite{Demirtas:2023fir}, conformal symmetry and defects \cite{Halverson:2024axc,Capuozzo:2025ozt,Robinson:2025ybg}, and renormalization \cite{Erbin:2022lls,Erbin:2021kqf,Sen:2025vzl} --- have been studied in this framework.

It is natural to wonder whether the NN-FT paradigm can accommodate fields with higher spin, such as a spin-$1$ gauge field $A_\mu$, which transform more non-trivially under (Euclidean) spacetime symmetries. In this Letter, we will answer this question in the affirmative for the example of Maxwell theory in four Euclidean dimensions. To do this, we must engineer an architecture $A^\mu_\theta$ and parameter density $P ( \theta )$ which satisfy three necessary properties:
\begin{enumerate}[label = (\roman*)]
    \item\label{correlators} The parameter-space integrals (\ref{scalar_correlator}) must reproduce the Maxwell correlation functions.

    \item\label{vector} $A^\mu_\theta$ must transform as a spacetime vector.

    \item\label{gauge} The theory must be gauge-invariant.
\end{enumerate}
Since Maxwell theory is free, one might expect that \ref{correlators} can be achieved by using a vector version of the cos-net architecture \cite{Halverson:2021aot} that realizes the free scalar (see \cite{Zhang:2026tss} for an optimality study among related architectures). This free scalar NN-FT has parameters which are all drawn independently. We will see that a version of this proposal succeeds, but with an important difference: unlike the scalar case, the parameters of the Maxwell architecture are \emph{not} all independent. Specifically, the vector of output weights must be chosen so that its spacetime direction is conditional upon that of the vector of input weights.

This simple modification is also closely related to \ref{vector} and \ref{gauge}. Physically, this is because the output weights control the polarization vector $\epsilon^\mu$ while the input weights are related to the momentum $k^\mu$. Both $\epsilon^\mu$ and $k^\mu$ are spacetime vectors. Whereas a rotation of $x^\mu$ in scalar NN-FT can be absorbed by a rotation of the input weights determining $k^\mu$ using the mechanism of \cite{Maiti:2021fpy}, if the output weights were drawn independently, this absorption could not enact a transformation of $\epsilon^\mu$ in Maxwell theory, contradicting \ref{vector}. Roughly speaking, conditioning the output weights on the input weights allows $\epsilon^\mu$ to transform, thus ensuring that $A^\mu$ behaves as a vector.

In Lorenz gauge, one has $k^\mu \epsilon_\mu = 0$, which is the statement that only the transverse polarizations of the photon are physical. Gauge transformations shift $\epsilon_\mu \to \epsilon_\mu + \beta k_\mu$ and must leave genuine observables unchanged. But as the output weights determine $\epsilon_\mu$, their distribution must ``know about'' the input weights that describe $k_\mu$ in order to ensure invariance under such gauge transformations. This again requires a conditional distribution. While independence breaking in NN-FT often leads to interactions, Maxwell NN-FT gives an interesting example where dependence between parameters preserves the Gaussian nature of the { infinite-width } theory (due to independence of neurons) but ensures appropriate behavior under rotations and gauge transformations.

NN-FT is both an analytical and a numerical toolkit. In the remainder of this Letter, we will elaborate upon the features summarized above -- proving analytically that our architecture and density obey \ref{correlators} - \ref{gauge} -- and present a continuum numerical simulation of Maxwell theory using NN-FT, reproducing the expected correlation functions via Monte Carlo parameter sampling.

\textbf{Review of NN-FT.}---A scalar neural network field theory is a pair 
$(\phi_\theta, P(\theta))$, where $\phi_\theta: \bR^d \to \bR$ is a neural network architecture with parameters $\theta$, and $P(\theta)$ is a density on the parameters. The associated parameter space partition function 
\begin{equation}
Z[J] = \bE_\theta[e^{\int d^dx \, J(x) \phi_\theta(x)}] = \int d\theta \, P(\theta) \, e^{\int d^dx \, J(x) \phi_\theta(x)}
\end{equation}
generates the  correlation functions (\ref{scalar_correlator}) by $J$-derivatives, enabling the study of the NN-FT even if the action is unknown. An example is the cos-net architecture
\begin{align}\label{scalar_architecture}
    \phi_\theta ( x ) = \frac{1}{\sqrt{N}} \sum_{i=1}^{N} a_i \cos \left( w_i \cdot x + b_i \right) \, .
\end{align}
Global symmetries may be ensured by the \emph{absorption mechanism} \cite{Maiti:2021fpy}, in which a transformation of the field is absorbed into a redefinition of parameters that leaves $Z[J]$ invariant. For example, consider a spacetime rotation $x \to R x$. Introduce the dummy variable $\tilde{w}_i = R^T w_i$. Then $\phi_\theta ( R x ) = \phi_{\tilde{\theta}} ( x )$, where $\tilde{\theta} = \{ a_i, \tilde{w}_i, b_i \}$. If the $w_i$ density is rotationally invariant, then the probability distributions for $P ( \theta )$ and $P ( \tilde{\theta} )$ coincide, and thus the induced distribution over configurations $\phi_\theta ( R x )$ agrees with that of $\phi_\theta ( x )$; it follows that all correlation functions enjoy rotation invariance. Thus, although individual field draws are not rotationally invariant, we say that the NN-FT is rotationally invariant \emph{in ensemble}. Similarly, if the construction is promoted to a vector field $\phi_\theta^j$ taking values in $\bR^M$ then the theory enjoys $O(M)$ symmetry if the density on output weights is $O(M)$ invariant.

\medskip
\textbf{Maxwell Fields from NN-FT.}---Gauge fields, on the other hand, are spacetime vector fields. Whereas a rotationally invariant $\phi_\theta^j$ theory enjoys symmetry under independent input and output rotations, Maxwell's theory of gauge fields with Lagrangian
\begin{align}
    \cL = \frac{g}{4} F_{\mu\nu}F^{\mu\nu} \, ,
\end{align}
where $F_{\mu\nu} = \partial_\mu A_\nu - \partial_\nu A_\mu$, involves a field $A_\mu$ transforming in the spin-$1$ representation of the spacetime symmetry group and thus must be invariant \emph{only} under simultaneous rotations of $x_\mu$ and $A_\mu$, not separate ones.

\medskip
We work on $\bR^4$ { with gauge coupling $e$ }and $ g = \frac{1}{e^2} >0$, gauge-fixing parameter $\xi\in[0,\infty)$, and UV cutoff $\Lambda>0$.  Like (\ref{scalar_architecture}), the network is formed from random Fourier features, a single hidden layer of width $N$, and $\cos$ activations:
\begin{equation}\label{eq:arch}
  \boxed{\;
  A^\mu(x)=\frac{1}{\sqrt{N}}\sum_{i=1}^{N}u_i^{\mu}\,
    \cos \bigl(w_i^\nu x_\nu + b_i\bigr) \, .
  \;}
\end{equation}
Here and henceforth, we typically suppress $\theta$ subscripts on architectures such as $A^\mu_\theta$ for brevity. The parameters are drawn independently across $i$ but not $\mu$,
\begin{align}\label{eq:measure}
  w_i^\mu &\sim \frac{d^4w}{\pi^2\Lambda^2|w|^2}\Big|_{B_\Lambda},
  \qquad
  b_i \sim \mathrm{Uniform}[0,2\pi),\notag\\
  u_i^\mu \,| \, w_i^\nu &\sim
    \cN\!\bigl(0,\,\sigma^2[\,P_\perp(w_i)+\xi\,\hat w_i\hat w_i^{\!\top}\,]^{\mu \nu} \bigr),
\end{align}
with output-weight scale $\sigma^2=\Lambda^2/(8\pi^2 g)$.  Here
$B_\Lambda=\{|w|\le\Lambda\}$ is the solid ball of radius $\Lambda$, $\hat w=w/|w|$, and
$P_\perp(w)=I_4-\hat w\hat w^{\!\top}$ projects onto the $3$-plane
orthogonal to $w$. Crucially, we have correlated the $\mu$-dependence of input and output weights by making the distribution of $u_i^\mu$ conditional on $w_i^\mu$, as motivated in the introduction.

\medskip
\textbf{Correlation functions.}---We now verify property \ref{correlators}. Since Maxwell theory is free, and likewise the architecture (\ref{eq:arch}) is Gaussian at infinite $N$ by central limit theorem arguments, it suffices to compute the two-point function, as this determines all $n$-point functions.

Each ingredient in the architecture and density is fixed by one feature of the target
propagator. The cosine and uniform bias $b$ give translation invariance and a
momentum-space $\delta$-function; the $|w|^{-2}$ density gives the
massless $1/k^2$ pole; $\Lambda$ is the sharp cutoff and the
radial measure ensures $SO(4)$ invariance; the conditional covariance supplies
the polarization tensor $\delta^{\mu\nu}+(\xi-1)\hat w^\mu\hat w^\nu$; and
$\sigma^2$ with the $1/\sqrt N$ scaling fix the residue to $1 / g$.
Equivalently, given $w_i$, the output weight $u_i$ is a sum of a transverse vector and a longitudinal scalar,
 \begin{gather}
 u_i^\mu = \tensor{P_\perp(w_i)}{^\mu_\nu} v_i^\nu +\,\ell_i\hat w_i^\mu \, , \\ 
 v_i^\mu \sim\cN(0,\sigma^2 I_4), \quad \ell_i\sim\cN(0,\sigma^2 \xi).
 \end{gather}
We compute the propagator directly in the neural network parameter space in Appendix~\ref{app:twopt}.  The computation involves a double-sum, but the average over the bias $b$ gives $\langle
A^\mu\rangle=0$ and kills the off-diagonal ($i\neq j$) neuron terms,
leaving a single-neuron expectation. Conditioning on $w$, the $u$-average
is the covariance and the bias average gives $\tfrac12\cos(w\!\cdot\!z)$;
the frequency integral localizes at $w=\pm k$ and evaluates the density at
$|w|=|k|$, yielding
\begin{equation}\label{eq:propk}
  \boxed{\;
   \widetilde C^{\mu\nu}_A (k)=\frac{1}{g\,k^2}
    \Bigl[\delta^{\mu\nu}-(1-\xi)\frac{k^\mu k^\nu}{k^2}\Bigr],
  \quad |k|\le\Lambda,
  \;}
\end{equation}
and zero for $|k|>\Lambda$. This is the covariant-gauge Maxwell propagator with gauge coupling $e$ and gauge-fixing parameter $\xi$. 
Inverting~\eqref{eq:propk} gives the kinetic operator
$\left( \widetilde C^{\mu\nu}_A (k) \right)^{-1}=g[\,k^2\delta^{\mu\nu}-(1-1/\xi)k^\mu k^\nu\,]$, the
momentum-space form 
of the integrand in
$S=\int d^4x\,[\tfrac{g}{4}F_{\mu\nu}F^{\mu\nu}+\tfrac{g}{2\xi}(\partial_\mu
A^\mu)^2]$. As $N\to \infty$ the measure on $A$ becomes a Gaussian process with this covariance, recovering free Maxwell theory exactly for all $|k|\le\Lambda$. As the UV cutoff $\Lambda$ is taken to infinity, one obtains Maxwell theory for all momenta.

\medskip
\textbf{Transformation under rotations.}---Next we check the transformation property \ref{vector}. The argument closely mirrors the one for the scalar field below equation (\ref{scalar_architecture}). 

Suppose that we again enact a spacetime rotation $x_\mu \to \tilde{x}_\mu = \tensor{R}{_\mu^\nu} x_\nu$ and make the parameter redefinitions $\tilde{w}_i^\mu = \tensor{R}{_\nu^\mu} w_i^\nu$, $\tilde{u}_i^\mu = \tensor{R}{_\nu^\mu} u^\nu_i$. Then one has the draw-by-draw identity
\begin{align}\label{draw_by_draw_A}
    A_\theta^\mu ( R x ) = \tensor{R}{^\mu_\nu} A^\nu_{\tilde{\theta}} ( x ) \, , \; \tilde{\theta} = \{ \tilde{u}_i^\mu , \tilde{w}_i^\mu , b_i \} \, ,
\end{align}
and we temporarily restore parameter subscripts for clarity. As in the scalar case, this identity follows from orthogonality,  since $w_i^\mu \tilde{x}_\mu = w_i^\mu \tensor{R}{_\mu^\nu} x_\nu = \tilde{w}_i^\mu x_\mu$.

We must still show invariance of the theory, which requires invariance of the partition function $Z[J_\mu(x)]$ under $J_\mu(x) \mapsto J^R_\mu(x) = R_\mu{}^\alpha J_\alpha(R^{-1}x)$. This may be absorbed into a transformation of the fields, requiring an invariance of the measure governing the fields. Ordinarily this is the Feynman path integral measure, but in NN-FT it is the measure on parameters.
Because the distribution of $(w_i^\mu, u_i^\mu)$ is rotationally invariant 
and the Jacobian for this change of variables is unity, this establishes that $\tilde{\theta}$ has the same probability distribution as $\theta$, and thus the same for $A^\mu_\theta(x)$ and $A^\mu_{\tilde{\theta}}(x) = \tensor{R}{_\nu^\mu} A^\nu_\theta(R x)$ as random fields. Invoking this while applying Eq.~\eqref{draw_by_draw_A} to every insertion therefore gives
\begin{align}
&\left\langle A^{\mu_1}(Rx_1)\cdots A^{\mu_n}(Rx_n)\right\rangle\notag\\
&\quad=R^{\mu_1}{}_{\nu_1}\cdots R^{\mu_n}{}_{\nu_n}
\left\langle A^{\nu_1}(x_1)\cdots A^{\nu_n}(x_n)\right\rangle.
\label{eq:vector-covariance}
\end{align}
This is precisely vector covariance in ensemble. A fixed draw need not be covariant by itself: its rotation is represented by another, equally distributed point in parameter space. This is global symmetry. The argument relied on the form of the conditional distribution of $u_i$ given $w_i$, which supplies the desired transformation property.

\medskip

\textbf{Gauge invariance.}---Finally, we turn to the more subtle issue \ref{gauge} of gauge invariance. In this section, we first establish finite-$N$ gauge invariance and then show how familiar gauge invariance is recovered as $N \rightarrow \infty$, in the $\xi \rightarrow \infty$ limit where gauge fixing is removed.

There are  three logically distinct issues: invariance of observables, closure of the NN configuration space, and invariance of the probability distribution. First, of course, the identity
\begin{equation}
 F_{\mu\nu}[A+\partial\lambda]=F_{\mu\nu}[A]
\end{equation}
holds for any sufficiently regular (or distributional) $\lambda$. This implies that all field strength correlation functions computed using the NN-Maxwell theory are gauge invariant. However, it does not guarantee that $A+\partial\lambda$ remains in the image of the finite-width architecture or that the parameter density is invariant, which we must show if we wish to exhibit a form of finite-$N$ gauge invariance. 

\medskip
\emph{Necessary Condition: Function Class Invariance}. Recall that as universal approximators, finite neural networks cannot exactly represent every continuous function. Therefore, under the transformation $A_\mu \mapsto A_\mu + \partial_\mu \lambda =: A'_\mu$, gauge invariance demands that $A$ and $A'$ realize the same set of functions. A simple choice that is sufficient to achieve this is that  $A'$ has the same architecture as $A$. This occurs when 
\begin{equation}\label{eq:finite-gauge-function}
 \lambda_{\{w,b\}}(x)=\frac{1}{\sqrt N}\sum_{i=1}^N c_i
 \sin(w_i\!\cdot\!x+b_i),\qquad c_i\in\mathbb R \, ,
\end{equation}
or in momentum space,
\begin{align}
 \widetilde\lambda_{\{w,b\}}(k)
 ={}&\frac{(2\pi)^4}{2i\sqrt N}\sum_{i=1}^N c_i
 \bigl[e^{ib_i}\delta^{(4)}(k-w_i)\notag\\[-1mm]
 &\hspace{24mm}-e^{-ib_i}\delta^{(4)}(k+w_i)\bigr].
\end{align}
With this gauge transformation, the transformed field is
\begin{equation}\label{eq:arch_prime}
  A^{\prime\mu } (x)=\frac{1}{\sqrt{N}}\sum_{i=1}^{N}\left(u_i^{\mu} + c_i \, w_i^\mu\right)\,
    \cos \bigl(w_i \cdot x+b_i\bigr) \, .
\end{equation}
This transformation is absorbed into parameter redefinitions as 
\begin{align}\label{eq:gauge-absorption}
 u_i^\mu \mapsto  u_i^\mu+c_iw_i^\mu \, , \; \ell_i \mapsto \ell_i+c_i|w_i| \, , \; \ell_i\equiv\hat w_i\!\cdot u_i,
\end{align}
where we see that the longitudinal component $\ell$ is what shifts under the gauge transformation.

\smallskip
The gauge transformation deserves comment.
Crucially, \eqref{eq:finite-gauge-function} had to re-use the same random variables $w_i^\mu$ and $b_i$ from the architecture \eqref{eq:arch} to make absorption (closure) possible.  In numerical simulations, this means the gauge transformation is draw-dependent. Despite inheriting some random variables from $A$, we emphasize the transformation amplitudes $c_i$ are independent of the field amplitudes $u_i^\mu$ appearing in $A$, giving the ability to change $\lambda$ independent of $A$.  
For { suitable } choices of $c_i$ and fixed parameter draws, as $N\to\infty$, $\lambda_{\{w,b\}}(x)$ converges to a tempered distribution $\lambda$ on $\bR^4$, i.e., { converges to the appropriate object to define a gauge transformation.}
Standard gauge invariance is defined in terms of an arbitrary, specified function or distribution $\lambda$ as a desired gauge transformation.
To recover it, note that the sampled pairs $(w_i,b_i)$ are { almost
surely dense in the support of their joint distribution at infinite
width, and coefficients $c_{i,N}=c_{i,N}(\{w,b\})$ may be chosen
so that $\lambda_{\{w,b\}}$ converges in the space of tempered distributions to any
prescribed target $\lambda$ compatible with the momentum cutoff. An arbitrary gauge transformation can then be achieved by taking $\Lambda \to \infty$.}
Further details and proofs are provided in Appendix \ref{app:cosine-completeness}.

\medskip
\emph{Sufficient Condition: Density Invariance}.  Absorption of the gauge transformation into the architecture is not sufficient to ensure invariance of the measure, which we now show. Conditioned on $w_i$, one has $u_i=u_i^\perp+\ell_i\hat w_i$ with $\ell_i\sim\mathcal N(0,\sigma^2\xi)$. This follows from diagonalizing the covariance matrix in (\ref{eq:measure}) using a basis whose first vector is $\hat{w}_i$ and whose remaining basis elements are orthogonal to $\hat{w}_i$; the covariance matrix becomes $\mathrm{diag} \left( \sigma^2 \xi, \sigma^2, \sigma^2, \sigma^2 \right)$, and the variance of $\ell_i = \hat{w}_i \cdot u_i$ is precisely that of the first basis element. At $\xi \geq 0$, the effect of the gauge transformation Eq.~\eqref{eq:gauge-absorption} only translates this Gaussian and changes the gauge-fixed density. At $\xi< \infty$ the $\ell$-density is not invariant under shifts; the gauge fixing breaks the gauge symmetry. On the other hand, the gauge fixing term is removed in the $\xi \to \infty$ limit, in which case we see that the $\ell$-density is invariant, but not normalizable.  Nevertheless, for every $N$ and finite $\xi$, functionals of $F$ are exactly invariant and $\xi$-independent because they involve only the transverse weights. Again, the conditional distribution of $u_i$ on $w_i$ plays a key role in this argument.

\begin{figure*}[t]
  \centering
  \includegraphics[width=\textwidth]{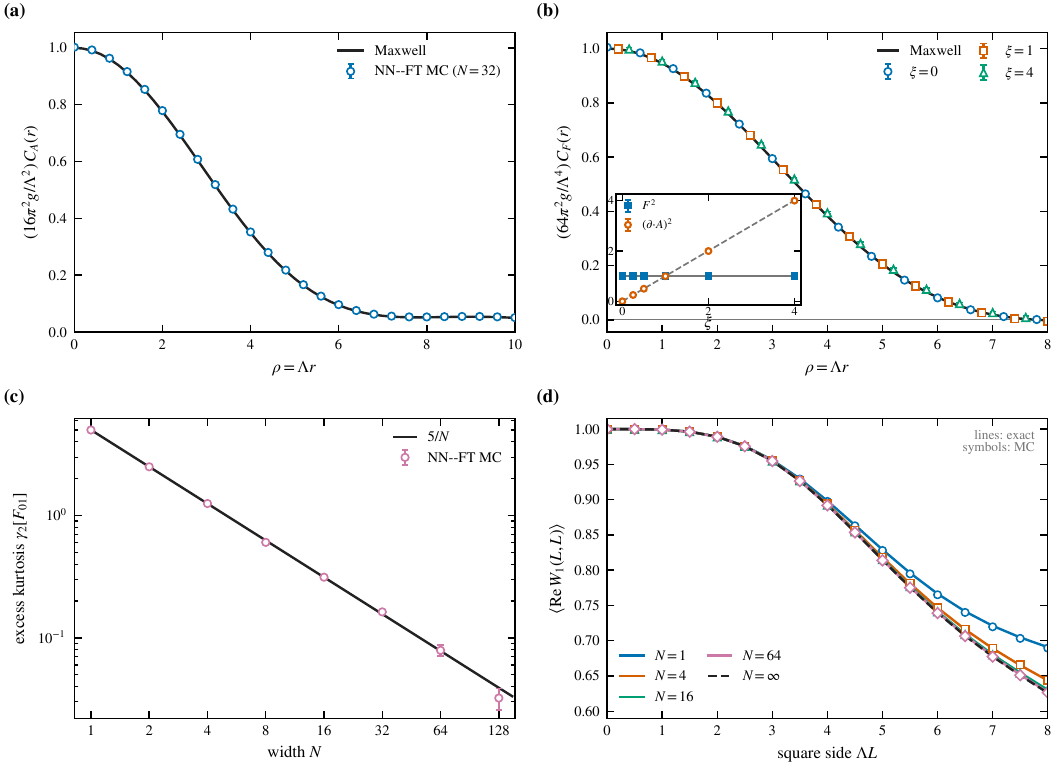}
  \caption{\textbf{Continuum Monte Carlo tests of the Maxwell NN--FT.}
  (a) Feynman-gauge position-space propagator.
  (b) Gauge-independent field-strength correlator for three gauge
  parameters; the inset contrasts the $\xi$-independent field-strength
  variance with the longitudinal variance proportional to $\xi$.
  (c) Gauge-invariant excess kurtosis, displaying the exact $5/N$
  finite-width correction.
  { (d) Square Wilson loops $W_q$ at probe charge $q=1$, at finite width and in the Gaussian-process limit. Curves denote the analytical predictions described in the text, and symbols denote independent field-draw estimates.}  Error bars show one standard
  error and are often smaller than the symbols.}
  \label{fig:maxwell-numerics}
\end{figure*}

\medskip
\textbf{Numerics.}---We sample the parameter density directly, without a
spacetime lattice or a Markov chain.  To obtain the power-law $w$-draws, note that $d^4 w = r^3 \, dr \, d \Omega_3$ in spherical coordinates, so the frequency measure in (\ref{eq:measure}) has radial density $\frac{2 r}{\Lambda^2}$ and is uniformly distributed on $S^3$. Therefore, to sample $w_i$, it suffices to draw $r=\Lambda\sqrt{U}$ with $U\sim{\rm Uniform}[0,1]$, draw $n=z/|z|\in S^3$ from $z\sim\mathcal N(0,I_4)$, and set $w=rn$. The conditional output weight is generated from an independent
$v\sim\mathcal N(0,I_4)$ as
$u=\sigma ( v_T+\sqrt{\xi}\,t n )$, 
where
$v_T=v-(v\cdot n)n$ and $t\sim\mathcal N(0,1)$. This implements the diagonalization of the covariance matrix discussed above, rescaling the independent Gaussian components so that each transverse component has variance $\sigma^2$ and the longitudinal component has variance $\sigma^2 \xi$. Unless noted otherwise,
we use $g=\Lambda=1$.  All uncertainties below are one-standard-error
estimates over $M$ independent, complete network draws.  We evaluate derivatives
of the network and line integrals of each cosine feature analytically.  
As an independent check, we also recomputed $F_{\mu \nu}$ using numerical differentiation and Wilson loops via quadrature on $768$ draws across various values of parameters like $\xi = 0, 1, 4$; the resulting observable means shifted by less than $0.002$ Monte Carlo standard errors, consistent with expected discretization errors.
Throughout this section, if $F_{\mu\nu}$ (potentially with fixed $\mu\nu$ values) is referred to inside a correlator without an argument, coincident spacetime points are to be assumed.

Figure~\ref{fig:maxwell-numerics}(a) compares the component-averaged
Feynman-gauge correlator $C_A(r)$ at (dimensionless) separation $\rho=\Lambda r$ to 
the exact expression 
\begin{equation}
 \frac{16\pi^2g}{\Lambda^2}C_A \left( \frac{\rho}{\Lambda} \right)
 =\frac{4[1-J_0(\rho)]}{\rho^2},
\end{equation}
where the $\rho=0$ value is defined by continuity.  The Monte Carlo
estimate is for width $N=32$ and $M=8\times10^4$ draws. 
Its largest pointwise
standardized residual is $1.46$, and its root-mean-square standardized
residual over the plotted separations is $1.10$. 
This agreement is not an
$N\to\infty$ effect: the two-point function is exact at every
width.

We next measure the gauge-invariant correlator
$C_F(r)=\langle F_{01}(0)F_{01}(r e_3)\rangle$, where $e_3$ is the unit vector along the $x^3$ direction.  Its cutoff Maxwell value is 
\begin{equation}
 C_F \left( \frac{\rho}{\Lambda} \right) =\frac{\Lambda^4}{2\pi^2g}
 \left[\frac{2[1-J_0(\rho)]}{\rho^4}
       -\frac{J_1(\rho)}{\rho^3}\right],
\label{eq:field-strength-position}
\end{equation}
with $C_F(0)=\Lambda^4/(64\pi^2g)$.  Figure~\ref{fig:maxwell-numerics}(b)
shows independent ensembles at $\xi=0,1,4$, using $N=32$ and
$M=6\times10^4$ for each value; all three agree with
Eq.~\eqref{eq:field-strength-position}.  The inset gives a separate local
check using $M=8\times10^4$:
\begin{equation}
 \frac{16\pi^2g}{3\Lambda^4}\langle F_{\mu\nu}F_{\mu\nu}\rangle=1,
 \qquad
 \frac{32\pi^2g}{\Lambda^4}
 \langle(\partial_\mu A_\mu)^2\rangle=\xi .
\end{equation}
For example, at $\xi=4$ the two estimates are
$0.9977(22)$ and $3.992(20)$, respectively.  In Landau gauge the measured
divergence is zero to floating-point precision.

Finite width first appears in connected correlators.  A particularly clean,
gauge-invariant diagnostic is the excess kurtosis
$\gamma_2[F_{01}]:=\langle F_{01}^4\rangle/
\langle F_{01}^2\rangle^2-3$.  The exact finite-$N$ result is $\gamma_2[F_{01}]=\frac{5}{N}$.
Panel (c) shows measurements from $M=5\times10^5$ draws at each width, where errors represent uncertainties across $50$ { batches}.  Representative values are
$4.994(60)$, $0.3130(83)$, and $0.0322(64)$ at
$N=1,16,128$, compared with $5$, $0.3125$, and $0.0391$.  This
exhibits the $O(1/N)$ non-Gaussian correction and its disappearance in the Gaussian process limit.

For a centered square in the $(0,1)$ plane, integrating the Gaussian
output weights and uniform bias gives the exact analytical prediction
for the finite-width NN-FT, 
\begin{align}
 \langle W_q\rangle_N
 &=\left\{\bE_w[e^{-x}I_0(x)]\right\}^{N},
 \qquad x=\frac{q^2H(w,L)}{4N},\nonumber\\
 H(w,L)&=\sigma^2L^4(w_0^2+w_1^2)
 \prod_{a=0}^{1}
 \left[\frac{\sin(w_aL/2)}{w_aL/2}\right]^2 ,
\end{align}
where $I_0$ is the modified Bessel function and $\bE_w$ denotes
expectation over the frequency distribution in Eq.~\eqref{eq:measure}.

In Maxwell theory, the loop flux $\Phi=\oint A_\mu dx^\mu$ is Gaussian,
so the Wilson-loop expectation is exactly
$\langle W_q\rangle=\exp[-q^2\langle\Phi^2\rangle/2]$.
Evaluating the flux variance for the square with the same sharp
momentum cutoff gives
\begin{align}
 \log\langle W_q\rangle_{\mathrm{Maxwell},\Lambda}
 &=-4q^2\int_0^L ds\,(L-s)\nonumber\\
 &\quad\times
 \left[C_A(s)-C_A\!\left(\sqrt{s^2+L^2}\right)\right]\nonumber\\
 &=-\frac{q^2}{4}\bE_w H(w,L),
\end{align}
where $C_A(r)=[1-J_0(\Lambda r)]/(4\pi^2gr^2)$, with its value
at the origin defined by continuity. Thus the exact cutoff Maxwell
result coincides with the $N\to\infty$ limit of the NN-FT prediction.
At finite width, the flux has the same variance but nonzero higher
cumulants, which modify the Wilson-loop expectation.

In panel (d), the solid curves are numerical evaluations of the
analytical finite-$N$ NN-FT predictions, while the dashed curve labeled
$N=\infty$ is the exact cutoff Maxwell result. The symbols are direct
averages of $\cos(q\Phi)$ over $M=1.5\times10^5$ independent, complete
network draws, with $\Phi$ evaluated analytically feature by feature.
The comparison therefore tests both the finite-width predictions and
their convergence toward Maxwell theory. At $\Lambda L=8$, for example,
the simulations give $0.6895(12)$ at $N=1$ and $0.6265(11)$ at $N=64$,
while the exact cutoff Maxwell value is $0.62598$. The measured
imaginary parts vanish within statistical errors. Since the
closed-loop current is transverse, the Wilson-loop expectation is
independent of $\xi$ at every $N$.

\medskip
\textbf{Discussion.}---In this work we have provided the first realization of gauge fields in NN-FT. Specifically, we have realized Maxwell theory as a NN-FT; see \cite{Dogra:2026hfa} for Maxwell CFT realized using the embedding formalism. As we emphasized, the key properties \ref{correlators} - \ref{gauge} are achieved by correlating the input and output weights in the vector cos-net architecture (\ref{eq:arch}). We argue that this architecture, which provably obeys the three stated necessary conditions, is also sufficient for representing the entire Maxwell theory in the infinite-width limit. It reproduces the Maxwell two-point function, and thus all higher-point functions by Gaussianity. This claim is also supported by independent numerical evidence which simulates both correlation functions of local operators and expectation values of non-local operators such as square Wilson loops.

There remain several interesting directions for future research. Perhaps the
most obvious is to generalize this construction to non-Abelian gauge theory
and perform analogous simulations in Yang--Mills theory. This would require
introducing an additional color index on the gauge field,
$A_\mu=A_\mu^aT_a$, and modifying the parameter density to encode interactions between gluons. More importantly, an infinitesimal
non-Abelian gauge transformation takes the nonlinear form $\delta_\lambda A_\mu^a
 =\partial_\mu\lambda^a+f^a{}_{bc}A_\mu^b\lambda^c$. The second term mixes color components and, in a Fourier-feature
architecture, also mixes momenta. Consequently, the simple shift of the
longitudinal output weights used in the Abelian theory will not close on a
generic finite-width cos-net. A non-Abelian construction may therefore
require an enlarged architecture which is closed under this mode mixing, or
else a notion of gauge invariance which emerges only after taking an
appropriate infinite-$N$ closure. The known NN-FT
realizations of ghost systems \cite{Frank:2026bui,Ferko:2026kkm} may also be useful for incorporating the
Faddeev-Popov sector in a gauge-fixed formulation.

Another natural direction is to couple the Maxwell gauge field $A^\mu$, or its non-Abelian generalization, to charged matter. Here the appropriate parameter density would not factorize into independent gauge and matter sectors: the correlations
between their parameters would encode vertices such as
$A_\mu\overline{\psi}\gamma^\mu\psi$, or their scalar-QED analogues. Gauge
invariance would require a simultaneous parameter-space absorption
mechanism implementing $A_\mu\longmapsto A_\mu+\partial_\mu\lambda$, $\psi\longmapsto e^{iq\lambda}\psi$,
or the corresponding transformation for a charged scalar. As
multiplication by a position-dependent phase generally mixes Fourier
features, closure at finite width is again nontrivial.

The finite-width theory may also be interesting in its own right, rather
than merely as an approximation to the Gaussian-process limit. At every
width it has exactly the Maxwell two-point function, while its connected
four-point functions and non-local observables receive calculable
finite-width corrections beginning at order $1/N$. It would be useful to
derive the effective action, if one exists, which governs the transverse
sector at finite $N$, and to investigate its properties as a deformation of free Maxwell theory. From this viewpoint, inverse width would serve as a controlled deformation parameter rather than merely as a measure of numerical approximation error.

Finally, it would be very interesting to extend this construction from
spin-$1$ to spin-$2$. A natural first step would be linearized gravity,
using a symmetric-tensor-valued network $h_{\mu\nu}(x)$ whose output-weight
distribution is conditioned on the input momenta. The analogue of the Abelian absorption
mechanism would implement the linearized diffeomorphism $h_{\mu\nu}\longmapsto h_{\mu\nu} + \partial_\mu\xi_\nu+\partial_\nu\xi_\mu$, while correlation functions of the linearized Riemann tensor would provide
gauge-invariant observables analogous to the field-strength correlators
studied here. One could then ask whether non-Gaussian correlations in
parameter space can reproduce the nonlinear graviton interactions arising
from the Einstein-Hilbert action and, ultimately, nonlinear
diffeomorphism invariance. Such a construction could provide a new way to study quantum gravity, both analytically and numerically.

\medskip
\noindent \textbf{Acknowledgements.---}We thank Ning Bao, Manas Dogra, Samuel Frank, Thomas Harvey, and Joydeep Naskar for helpful discussions. J.H. is supported in part, and C.F. is supported, by the NSF under Cooperative Agreement PHY-2019786, the NSF AI Institute for Artificial Intelligence and Fundamental Interactions.
We thank the Institute for Nuclear Theory at the University of Washington for its kind hospitality and stimulating research environment. This research was supported in part by the INT's U.S. Department of Energy grant No.\ DE-FG02-00ER41132.

\appendix
\setcounter{secnumdepth}{1}
\onecolumngrid
\section{Explicit two-point function calculation}

\label{app:twopt}

This appendix derives the propagator~\eqref{eq:propk} from the
architecture~\eqref{eq:arch}--\eqref{eq:measure}. The computation simplifies once one conditions on the
frequency $w$: given $w$, the bias $b$ and the output weight $u$ are
independent, so their averages factor.  We use the conventions $\widetilde
f(k)=\int d^4z\,e^{-ik\cdot z}f(z)$ and $\int d^4z\,e^{iq\cdot
z}=(2\pi)^4\delta^{(4)}(q)$, and the parameter densities given in (\ref{eq:measure}), namely
$w\sim p(w)=1/(\pi^2\Lambda^2|w|^2)$ on $B_\Lambda$,
$u\,|\,w\sim\cN(0,\Sigma(w))$ with
$\Sigma^{\mu\nu}(w)=\sigma^2[P_\perp^{\mu\nu}(w)+\xi\,\hat w^\mu\hat w^\nu]$,
$\sigma^2=\Lambda^2/(8\pi^2 g)$, and $b\sim\mathrm{Uniform}[0,2\pi)$.

It is convenient to first compute the one-point function. Conditioning on $w_i$, the weight $u_i\,|\,w_i$
is zero-mean and independent of $b_i$, so
$\bE[u_i^\mu\cos(w_i\!\cdot\!x+b_i)]=\bE_{w_i,b_i}[\bE[u_i^\mu|w_i]\cos(\cdots)]=0$. Thus $\langle A^\mu(x)\rangle=0$ for every $N$, as expected.

Next we consider the two-point function, which we denote by $C^{\mu \nu}_A ( x, y )$ to distinguish it from the scalar Feynman gauge quantity $C_A ( r )$ discussed in the numerics section. The two-point function takes the form 
\begin{align}
    C^{\mu\nu}_A (x,y)=\tfrac1N\sum_{i,j}\langle u_i^\mu u_j^\nu\cos(w_i\!\cdot\!x+b_i)\cos(w_j\!\cdot\!y+b_j)\rangle \, .
\end{align}
The expectation over $b$ causes each term with $i \neq j$ to vanish. Since the neurons are i.i.d.\ and zero-mean, the remaining $N$ diagonal terms are identical, so the two-point function reduces to
\begin{equation}
  C^{\mu\nu}_A (x,y)=\bigl\langle u^\mu u^\nu\,
    \cos(w\!\cdot\!x+b)\cos(w\!\cdot\!y+b)\bigr\rangle .
\end{equation}
As in the one-point function calculation, we now condition on $w$. Given $w$, the $u$-average
is the covariance itself,
\begin{equation}
  \bE[u^\mu u^\nu\,|\,w]=\Sigma^{\mu\nu}(w)
  =\sigma^2\bigl[\delta^{\mu\nu}+(\xi-1)\hat w^\mu\hat w^\nu\bigr]
\end{equation}
The bias average, by the product-to-sum identity with the $2b$ term
averaging to zero, is $\bE_b[\cos(w\!\cdot\!x+b)\cos(w\!\cdot\!y+b)]
=\tfrac12\cos(w\!\cdot\!z)$, $z\equiv x-y$.  Hence
\begin{equation}\label{two_point_function_intermediate}
  C^{\mu\nu}_A (z)=\frac{\sigma^2}{2}\int_{B_\Lambda}p(w)\,d^4w\;
    \bigl[\delta^{\mu\nu}+(\xi-1)\hat w^\mu\hat w^\nu\bigr]\cos(w\!\cdot\!z).
\end{equation}
It is most transparent to express (\ref{two_point_function_intermediate}) in momentum space. Using the Fourier transform $\int
d^4z\,e^{-ik\cdot z}\cos(w\!\cdot\!z)=\tfrac12(2\pi)^4[\delta^{(4)}(k-w)
+\delta^{(4)}(k+w)]$, the $w$-integral localizes at $w=\pm k$.  For
$|k|\le\Lambda$ both points lie in $B_\Lambda$, and since the integrand is
even in $w$ they contribute equally, evaluating the density at $w=k$,
$p(k)=1/(\pi^2\Lambda^2 k^2)$:
\begin{equation}
  \widetilde C^{\mu\nu}_A (k)
  =\underbrace{\frac{\sigma^2(2\pi)^4}{2}\,p(k)}_{=\,1/(g k^2)}
   \bigl[\delta^{\mu\nu}+(\xi-1)\hat k^\mu\hat k^\nu\bigr].
\end{equation}
Here we have used $\sigma^2 = \frac{\Lambda^2}{8 \pi^2 g}$ to express the constant prefactor in terms of $g$. With $\hat k^\mu\hat
k^\nu=k^\mu k^\nu/k^2$ and $(\xi-1)=-(1-\xi)$ this reproduces
Eq.~\eqref{eq:propk}.  As a check, $\xi=1$ gives the Feynman form
$\delta^{\mu\nu}/(g k^2)$, while $\xi=0$ gives the transverse (Landau) form
$(\delta^{\mu\nu}-k^\mu k^\nu/k^2)/(g k^2)$, which satisfies $k_\mu\widetilde
C^{\mu\nu}_A (k)=0$.

\section{Distributional completeness of the cosine architecture}
\label{app:cosine-completeness}

In this Appendix we wish to show that the cosine architecture in the main text converges to an arbitrary tempered distribution in the infinite width limit; i.e. it realizes the distribution space used in constructive field theory. The argument follows a version of Theorem~VII.4 in \cite{Poletti:2025distributions}. We include a separate proof to make this work self-contained.

Let $T\in\mathcal S'(\mathbb R^4)$ be a real tempered distribution, where
$\mathcal S(\mathbb R^4)$ is the Schwartz space of smooth functions whose
derivatives decay faster than any power at infinity. A standard result in
distribution theory is that one can choose a sequence
$H_m\in C_c^\infty(\mathbb R^4)$, where $C_c^\infty$ denotes smooth
compactly supported functions, such that
$H_m\longrightarrow \widetilde T$
distributionally, where $\tilde T$ is the Fourier transform of $T$; see, for example, \cite{doi:10.1142/5314}.
Define the corresponding sequence
\[
F_m(x)=\int \frac{d^4k}{(2\pi)^4}\,H_m(k)e^{ik\cdot x}.
\]
Then $F_m\to T$ distributionally, so that
\[
\int d^4x\,F_m(x)g(x)\longrightarrow T(g)
\qquad
\text{for every }g\in\mathcal S(\mathbb R^4).
\]
Since $T$ is real, we may choose
$H_m(-k)=\overline{H_m(k)}$, so that each $F_m$ is real.

For fixed $m$, compactness of $\operatorname{supp}H_m$ allows us to
partition it into finitely many non-overlapping small boxes $C_{mj}$.
Choose $k_{mj}\in C_{mj}$ and require each box to have diameter at most
$\delta_m$, meaning that
\[
|k-k_{mj}|\le \delta_m
\qquad
\text{for all }k\in C_{mj}.
\]
The contribution of the $j$th box to $F_m$ is
\[
F_{mj}(x)=
\int_{C_{mj}}\frac{d^4k}{(2\pi)^4}\,
H_m(k)e^{ik\cdot x},
\]
and define its integrated Fourier weight by
\[
A_{mj}=
\int_{C_{mj}}\frac{d^4k}{(2\pi)^4}\,H_m(k).
\]
Approximating the whole box by its representative momentum gives the finite
cosine sum in position space
\[
p_m(x)=
\sum_j |A_{mj}|
\cos\!\left(k_{mj}\cdot x+\arg A_{mj}\right),
\]
i.e.\ the Fourier integral defining $F_m$ is approximated by replacing each
small momentum-space cell by its total Fourier weight $A_{mj}$ concentrated
at the representative momentum $k_{mj}$.

We now bound the error. Using
$p_m(x)=\operatorname{Re}\sum_j A_{mj}e^{ik_{mj}\cdot x}$ and
$|\operatorname{Re}z|\le |z|$,
\begin{align}
|p_m(x)-F_m(x)|
&\le
\sum_j
\left|
\int_{C_{mj}}
\frac{d^4k}{(2\pi)^4}
H_m(k)
\left[
e^{ik_{mj}\cdot x}-e^{ik\cdot x}
\right]
\right|
\notag\\
&\le
\sum_j
\int_{C_{mj}}
\frac{d^4k}{(2\pi)^4}
\left|
H_m(k)
\left[
e^{ik_{mj}\cdot x}-e^{ik\cdot x}
\right]
\right|
\notag\\
&=
\sum_j
\int_{C_{mj}}
\frac{d^4k}{(2\pi)^4}
|H_m(k)|
\left|
e^{ik_{mj}\cdot x}-e^{ik\cdot x}
\right|.
\end{align}
Since $|k-k_{mj}|\le\delta_m$ inside each cell,
\[
\left|e^{ik_{mj}\cdot x}-e^{ik\cdot x}\right|
=
\left|e^{i(k_{mj}-k)\cdot x}-1\right|
\le
\left|(k_{mj}-k)\cdot x\right|
\le
|k_{mj}-k|\,|x|
\le
\delta_m |x|.
\]
where the first inequality uses $|e^{iu}-1|\le |u|$, the second uses
Cauchy--Schwarz, and the third uses $|k-k_{mj}|\le\delta_m$ inside each cell.
Therefore
\[
|p_m(x)-F_m(x)|
\le
\delta_m |x|
\int\frac{d^4k}{(2\pi)^4}|H_m(k)|
=
\delta_m M_m |x|,
\]
where
\[
M_m=
\int\frac{d^4k}{(2\pi)^4}|H_m(k)|.
\]
Thus, for every $g\in\mathcal S$,
\[
\left|
\int d^4x\,[p_m(x)-F_m(x)]g(x)
\right|
\le
\delta_mM_m
\int d^4x\,|x|\,|g(x)|.
\]
Choose the boxes sufficiently small that $\delta_mM_m\to0$. Then
$p_m-F_m\to0$ distributionally, and since $F_m\to T$, it follows that
$p_m\longrightarrow T$
distributionally.

\section{Comments on constructive field theory and numerical simulations}

We end with comments on interesting recent results in the literature relating NN-FT and constructive field theory. In \cite{Harvey:2026iqd}, a theorem of Harvey showed that, under appropriate assumptions, two-point functions in NN-FT cannot be reflection positive (RP); this is the constraint on Euclidean correlators related to unitarity of a Lorentzian continuation in the Osterwalder-Schrader (OS) construction \cite{Osterwalder:1973dx}. However, the free-scalar propagator is realized in many finite-width NN-FTs, e.g. \cite{Halverson:2021aot, Frank:2026bui, Zhang:2026tss}, including continuous families of realizations at finite-width \cite{Zhang:2026tss} that converge to the same neural network Gaussian process at infinite width. There is no contradiction, because these examples violate the finite pointwise variance assumption of \cite{Harvey:2026iqd}, a possibility already noted in that work; we note it for the reader who may not realize the importance of the assumption. On the other hand, \cite{Ferko:2026luj} also studies higher-point functions and shows that it is not possible to realize the full OS construction with finite-width NN-FT. Together, \cite{Harvey:2026iqd,Ferko:2026luj} are complementary works that demonstrate limitations of NN-FT simulations on a finite-precision classical computer.

It is interesting to ask more generally: what are the limitations of studying QFT on a finite-precision classical computer? The answer could be exemplified in the context of NN-FT, lattice field theory, or other approaches. For instance: finite-width NN-FT is a continuum theory that can exhibit full Euclidean symmetry, while not being able to simultaneously realize all of the other OS axioms \cite{Ferko:2026luj}, whereas lattice field theory at finite spacing breaks Euclidean symmetry and has a version of RP consistent with the broken symmetry; it is an open question whether finite-width NN-FT can realize the same amount of RP as in lattice field theory. These and other questions could be addressed by better understanding limitations of finite-precision simulations and the relation to constructive field theory.

More generally, it is interesting to remember that even the finite-cutoff free scalar is not RP, nor is effective field theory, in general. This does not preclude a RP continuum limit, as occurs for the free scalar, although the existence of such a limit cannot generally be inferred from the IR description alone.  

\bibliographystyle{utphys}
\bibliography{refs}

@inproceedings{Erbin:2022lls,
    author = "Erbin, Harold and Lahoche, Vincent and Samary, Dine Ousmane",
    title = "{Renormalization in the neural network-quantum field theory correspondence}",
    eprint = "2212.11811",
    archivePrefix = "arXiv",
    primaryClass = "hep-th",
    month = "12",
    year = "2022"
}

@article{Harvey:2026iqd,
    author = "Harvey, Thomas R.",
    title = "{What Neural Network Field Theory Can and Cannot Realise on a Computer}",
    eprint = "2608.21523",
    archivePrefix = "arXiv",
    primaryClass = "hep-th",
    month = "8",
    year = "2026"
}

@article{Ferko:2026luj,
    author = "Ferko, Christian and Mutchler, Aaron",
    title = "{Neural Network Field Theory at Finite Width}",
    eprint = "2608.21588",
    archivePrefix = "arXiv",
    primaryClass = "hep-th",
    month = "8",
    year = "2026"
}

@article{Dogra:2026hfa,
    author = "Dogra, Manas and Halverson, James and Naskar, Joydeep",
    title = "{Spinning Conformal Correlators from Neural Networks}",
    eprint = "2608.15001",
    archivePrefix = "arXiv",
    primaryClass = "hep-th",
    month = "8",
    year = "2026"
}

@article{Poletti:2025distributions,
  author = {Poletti, Mario and Martelli, Mario},
  title = {Continuous linear shift invariant systems on tempered
           distributions: convolutors and convolution},
  journal = {Bollettino dell'Unione Matematica Italiana},
  volume = {18},
  pages = {1009--1026},
  year = {2025},
  doi = {10.1007/s40574-024-00447-2}
}

@article{Ferko:2026kkm,
    author = "Ferko, Christian and Frank, Samuel and Halverson, James and Jejjala, Vishnu",
    title = "{Anomalies in neural network field theory}",
    eprint = "2605.12488",
    archivePrefix = "arXiv",
    primaryClass = "hep-th",
    doi = "10.1007/JHEP10(2026)143",
    journal = "JHEP",
    volume = "10",
    pages = "143",
    year = "2026"
}

@article{Robinson:2025ybg,
    author = "Robinson, Brandon",
    title = "{Virasoro Symmetry in Neural Network Field Theories}",
    eprint = "2512.24420",
    archivePrefix = "arXiv",
    primaryClass = "hep-th",
    month = "12",
    year = "2025"
}

@article{Erbin:2021kqf,
    author = "Erbin, Harold and Lahoche, Vincent and Samary, Dine Ousmane",
    title = "{Non-perturbative renormalization for the neural network-QFT correspondence}",
    eprint = "2108.01403",
    archivePrefix = "arXiv",
    primaryClass = "hep-th",
    reportNumber = "MIT-CTP/5309",
    doi = "10.1088/2632-2153/ac4f69",
    journal = "Mach. Learn. Sci. Tech.",
    volume = "3",
    number = "1",
    pages = "015027",
    year = "2022"
}

@article{Sen:2025vzl,
    author = "Sen, Srimoyee and Vaidya, Varun",
    title = "{Viability of perturbative expansion for quantum field theories on neurons}",
    eprint = "2508.03810",
    archivePrefix = "arXiv",
    primaryClass = "hep-th",
    doi = "10.1088/2632-2153/ae6d7d",
    journal = "Mach. Learn. Sci. Tech.",
    volume = "7",
    number = "3",
    pages = "035044",
    year = "2026"
}

@article{Capuozzo:2025ozt,
    author = "Capuozzo, Pietro and Robinson, Brandon and Suzzoni, Benjamin",
    title = "{Conformal defects in neural network field theories}",
    eprint = "2512.07946",
    archivePrefix = "arXiv",
    primaryClass = "hep-th",
    doi = "10.1007/JHEP05(2026)124",
    journal = "JHEP",
    volume = "05",
    pages = "124",
    year = "2026"
}

@article{Osterwalder:1973dx,
	title        = {{Axioms for Euclidean Green's functions}},
	author       = {Konrad Osterwalder and Robert Schrader},
	year         = 1973,
	journal      = {Communications in Mathematical Physics},
	publisher    = {Springer},
	volume       = 31,
	number       = 2,
	pages        = {83 -- 112}
}

@article{Halverson:2020trp,
    author = "Halverson, James and Maiti, Anindita and Stoner, Keegan",
    title = "{Neural Networks and Quantum Field Theory}",
    eprint = "2008.08601",
    archivePrefix = "arXiv",
    primaryClass = "cs.LG",
    doi = "10.1088/2632-2153/abeca3",
    journal = "Mach. Learn. Sci. Tech.",
    volume = "2",
    number = "3",
    pages = "035002",
    year = "2021"
}

@article{Ferko:2026ken,
    author = "Ferko, Christian and Halverson, James and Jejjala, Vishnu and Robinson, Brandon",
    title = "{Topological effects in neural network field theory}",
    eprint = "2604.02313",
    archivePrefix = "arXiv",
    primaryClass = "hep-th",
    doi = "10.1007/JHEP09(2026)232",
    journal = "JHEP",
    volume = "09",
    pages = "232",
    year = "2026"
}

@article{Zhang:2026tss,
    author = "Zhang, Zhengkang",
    title = "{Optimal architecture and fundamental bounds in neural network field theory}",
    eprint = "2604.27050",
    archivePrefix = "arXiv",
    primaryClass = "hep-th",
    doi = "10.1103/4fzf-tpsv",
    journal = "Phys. Rev. D",
    volume = "114",
    number = "6",
    pages = "L061704",
    year = "2026"
}

@article{Maiti:2021fpy,
  author        = "Maiti, Anindita and Stoner, Keegan and Halverson, James",
  title         = "{Symmetry-via-Duality: Invariant Neural Network Densities from Parameter-Space Correlators}",
  eprint        = "2106.00694",
  archivePrefix = "arXiv",
  primaryClass  = "cs.LG",
  year          = "2021"
}

@article{Halverson:2021aot,
  author        = "Halverson, James",
  title         = "{Building Quantum Field Theories Out of Neurons}",
  eprint        = "2112.04527",
  archivePrefix = "arXiv",
  primaryClass  = "hep-th",
  year          = "2021"
}

@article{Demirtas:2023fir,
  author        = "Demirtas, Mehmet and Halverson, James and Maiti, Anindita and Schwartz, Matthew D. and Stoner, Keegan",
  title         = "{Neural network field theories: non-Gaussianity, actions, and locality}",
  eprint        = "2307.03223",
  archivePrefix = "arXiv",
  primaryClass  = "hep-th",
  doi           = "10.1088/2632-2153/ad17d3",
  journal       = "Mach. Learn. Sci. Tech.",
  volume        = "5",
  number        = "1",
  pages         = "015002",
  year          = "2024"
}

@article{Halverson:2024axc,
  author        = "Halverson, James and Naskar, Joydeep and Tian, Jiahua",
  title         = "{Conformal fields from neural networks}",
  eprint        = "2409.12222",
  archivePrefix = "arXiv",
  primaryClass  = "hep-th",
  doi           = "10.1007/JHEP10(2025)039",
  journal       = "JHEP",
  volume        = "10",
  pages         = "039",
  year          = "2025"
}

@article{Frank:2025zuk,
  author        = "Frank, Samuel and Halverson, James and Maiti, Anindita and Ruehle, Fabian",
  title         = "{Fermions and Supersymmetry in Neural Network Field Theories}",
  eprint        = "2511.16741",
  archivePrefix = "arXiv",
  primaryClass  = "hep-th",
  year          = "2025"
}

@article{Frank:2026bui,
  author        = "Frank, Samuel and Halverson, James",
  title         = "{String Theory from Infinite Width Neural Networks}",
  eprint        = "2601.06249",
  archivePrefix = "arXiv",
  primaryClass  = "hep-th",
  year          = "2026"
}

@article{Ferko:2026axm,
  author        = "Ferko, Christian and Halverson, James and Mutchler, Aaron",
  title         = "{Universality of Neural Network Field Theory}",
  eprint        = "2601.14453",
  archivePrefix = "arXiv",
  primaryClass  = "hep-th",
  year          = "2026"
}

@book{doi:10.1142/5314,
author = {Strichartz, Robert S},
title = {A Guide to Distribution Theory and Fourier Transforms},
publisher = {WORLD SCIENTIFIC},
year = {2003},
doi = {10.1142/5314},
address = {},
edition   = {},
URL = {https://www.worldscientific.com/doi/abs/10.1142/5314},
eprint = {https://www.worldscientific.com/doi/pdf/10.1142/5314}
}

\end{document}